# Tunable band gap of graphane nanoribbons under uniaxial elastic strain: a first-principles study

Yan Zhang [1], Xiaojun Wu [2,†] Qunxiang Li [1], and Jinlong Yang [1,†]

[1]*Hefei National Lab for Physical Sciences at the Microscale, University of Science and Technology of China, Anhui, Hefei, 23002, People's Republic of China*

[2]*CAS Key Lab of Materials for Energy Conversion, Hefei National Lab for Physical Science at the Microscale, and Department of Material Science and Engineering, University of Science and Technology of China, Anhui, Hefei, 23002, People's Republic of China*

In this Letter, we investigate the strain-induced band-gap modulation of both armchair and zigzag graphane nanoribbons based on the first-principles calculations. Within the elastic range, the band gap changes linearly with the uniaxial strain, where the band-gap of graphane nanoribbonsis more sensitive to compressive than tensile deformation. The band-gap deformation mainly originates from the shift of valence band edge of graphane nanoribbons under stain. Our results imply the great potential of graphane nanoribbons in the pressure sensor and optical electronics applications.

[†] To whom correspondence should be addressed.

E-mail: jlyang@ustc.edu.cn , xjwu@ustc.edu.cn

Graphene, true two-dimensional atomic layer material, has attracted considerable research attention for its intriguing physical properties and high potential for nanoelectronic applications [1-3]. Graphene is a semiconductor with a zero band gap. By cutting graphene into nanoribbons, the energy band gap of graphene can be tailored to meet the need of band-gap engineering in low-dimensional materials [4, 5]. Previous theoretical works have shown that the band gap of graphene nanoribbons can be effectively tuned by strain, but the band gap behavior largely depends on both their edge structures and widths [6-11]. For example, in the elastic deformation range, the energy band gap of zigzag graphene nanoribbons are less sensitive to the uniaxial strain, while the band gap of armchair graphene nanoribbons exhibits a sawtooth-shaped oscillation[6], blocking their applications.

Recently, fully hydrogenated graphene, namely graphane, has been first theoretically predicted by Sofo *et al.*[12], and then experimentally realized by exposing graphene to hydrogen plasma [13] or dissociating hydrogen silsequioxane on graphene [14]. Interesting properties of graphane such as markedly shorter lattice period than that of graphene, reversible hydrogenation with changing temperature[13,14], and a wide direct band gap[12,15], have been revealed. Extensive theoretical studies have also been performed on the graphane with defects[15,16], fluorine-substituted graphane[17], dehydrogenated graphane[18], and the superconductivity of doped graphane[19], *etc.*. In particular, graphane nanoribbons (GANRs) are wide direct band gap semiconductor regardless of their edge structure [20, 21], and the binding energy of hydrogen on graphene can be improved with biaxial strain[22]. The band-gap deformation for two-dimensional graphane is positive and higher than that of fully fluorine-substituted graphane [23]. The in-plane stiffness is smaller than those of graphene and the band gap can be strongly modified by applied strain in the elastic range [24]. Experimentally, it has been shown that GANRs can be obtained from hydrogenating and unzipping carbon nanotube [25].

In this letter, we use the first-principles method to systemically study the geometric, electronic and mechanical properties of both armchair and zigzag GANRs under uniaxial elastic strain. We observe a linear relationship between uniaxial strain and energy band gap for both zigzag and armchair GANRs. The band gap is more sensitive to compressing than tensile strain. This linear relationship between the band gap and uniaxial strain endows GANRs great potential in pressure sensor and optical electronics. To our knowledge, this is the first report on the band-gap modulation of GANRs under uniaxial elastic strain.

The first-principles calculations are performed using VASP code [26, 27]. The projector-augmented wave method [28, 29] for the electron-ion interaction, and the generalized gradient approximation of the PW91 functional [30] are used to describe electronic exchange and correlation. A kinetic-energy cutoff of 400eV is chosen for the plane wave basis set. Two adjacent nanoribbons are separated by a vacuum region of at least 13.0 Å. A K-point mesh of 1×1×7 grid was used for geometry optimization, and 1×1×17 grid for static electronic structures calculations. The axial lattice constant and atomic positions are fully relaxed until the force on each atom was less than 0.02 eV/ Å, and the convergence criteria for energy is $10^{-5}$ eV.

Fig. 1(a) shows the optimized structures of two examples of GANRs, 6-zigzag and 13-armchair GANRs. The corresponding width is 13.79 and 17.08 Å, respectively. In our calculations, the edge carbon atoms are saturated with hydrogen. The number $N$ in *N-zigzag or N-armchair* stands for the number of zigzag chains or dimer lines along the ribbon direction for a zigzag or armchair GANR, respectively [20,21]. In 13-armchair GANR, the optimized lengths are 1.54 and 1.52 Å for the inner and edge C-C bonds, respectively. In 6-zigzag GANR, the optimized inner and edge C-C bond lengths are 1.54 and 1.53 Å, respectively. The calculated band structures, as shown in Fig. 1(b), indicate that 6-zigzag and 13-armchair GANRs are nonmagnetic semiconductors with a direct band gap of 3.80 and 3.71 eV, respectively. As the GANRs' widths increase, their energy band gaps decreases monotonically, following the relationship of $E_{gap}$ = 3.47+1.71×$exp$(-N/6.76) and $E_{gap}$ = 3.49+2.18×$exp$(-N/3.14) for armchair and zigzag GANRs, respectively. The plotted charge density profiles of valence and conduction band edge (VBE and CBE) at the Γ point show that these bands mainly contributed by the inner part of GANRs [31]. These observations consist well with previous theoretical works [20, 21].

To modulate the energy band gap of GANR, a uniaxial elastic strain is applied along the ribbons' direction by stretching or compressing nanoribbon. The stress is defined as $\varepsilon = (c - c_0)/c_0$, where $c$ and $c_0$ are the periodic length of GANRs with and without deformation. Stretching or compressing ribbon corresponds to a positive or negative value of $\varepsilon$, respectively. In zigzag GANRs, the C-C bond perpendicular to strain direction does not changes obviously, while others are slightly elongated or shortened with tensile or compressing deformation. In armchair GANRs, all C-C bonds are slightly elongated or shortened with tensile or compressing deformation. The relationship between the energy band gap and uniaxial strain for 6-zigzag and 13-armchair GANRs are illustrated in Fig. 3(a). Clearly, the band gap can be effectively tuned by uniaxial elastic strain. The band gaps change remarkably from 2.49 to 4.11 eV and 2.04 to 4.21 eV for 13-armchair and 6-zigzag GANRs when $\varepsilon$ changes from -10.0% to 10.0%, respectively. Moreover, in the examined elastic range, the energy band gap of both armchair and zigzag GANRs changes almost linearly with the strain when stretching or compressing nanoribbons. Moreover, the energy band gap of GANR is more sensitive to the compressive than tensile strain. This monotonic behavior of energy band gap in GANR with elastic strain is quite different from the nonlinear behavior in GRNs, suggesting the potential applications of GANRs in band-gap engineering field, such as pressure sensor and tunable optical electronics at nanoscale.

It should be pointed out that the linear relationship of the band gap with the uniaxial stain is independent on nanoribbons' width. Taking the armchair GANRs as one example, Fig. 3(b) shows that the profiles energy band gap versus stress are almost parallel for armchair GANRs with different widths ($N$ ranges from 11 to 25).

To understand the band-gap modulation with the uniaxial elastic strain, we plotted the electronic band structures of 6-zigzag and 13-armchair GANRs under the strain of $\varepsilon = \pm 5\%$, as shown in Fig.4. The modification on the band gap with uniaxial strain mainly comes from the shift of VBE, e.g., the VBE of 13-armchair GANR shifts downward markedly from -3.31 to -4.02 eV, while CBE shifts upward slightly from -0.17 to -0.10 eV when $\varepsilon$ changes from -5% to 5%. The reason for this difference is that the charge density of VBE mainly locates on the C-C bonds which undergo deformation under stain, whereas the charge density of CBE mainly locates on the central atoms, as shown in Fig. 1. Similar behaviors can be found in 6-zigzag GANRs.

We also calculated the elastic constant of 6-zigzag and 13-armchair GANRs. For the one-dimensional GANRs, the Young's modulus can be calculated with $Y = (\partial^2 E/\partial \varepsilon^2)/V_0$, where $E$ is the total energy of GANR per unit cell. $V_0$ is the relaxed volume of GANR per unit cell without uniaxial strain, defined as $V_0 = w \times h_0 \times c_0$, where $w$ is the width, and $h_0$ is the thickness of GANR. Here, $h_0$ is chosen as 5.10 Å, equaling the vertical distance between two H atoms from both surfaces of graphane plus van de Waals distance between two H atoms (about 2.40 Å). The calculated Young's modulus of 6-zigzag and 13-armchair GANRs are about 230 and 220 GPa, respectively.

These values are quite smaller than those of graphene nanoribbons (about 1.01 TPa[32]).

In conclusion, we use the first-principles method to study the band-gap modulation of GANRs under uniaxial elastic strain. The band gap of GANRs has a almost linear relationship with uniaxial strain when stretching or compressing nanoribbons, and the band gap is more sensitive to compressive than tensile strain. The origin of band-gap modulation is the modification of VBE under uniaxial strain, where the charge density of VBE mainly locates on the deformed C-C bonds under strain. The calculated Young's modulus of GANR is quite smaller than that of graphene nanoribbon. Our studies indicate that GANRs have great potential in the pressure sensors and tunable optical electronics applications at nanoscale.

We thank Dr. Shuanglin Hu for the helpful discussion. This work was partially supported by the National Basic Research Program of China (Nos.2011CB921400, 2012CB922001), by the National Natural Science Foundation of China (Nos.20873129, 11004180, 91021004, 11074235, 11034006), by the Fundamental Research Funds for the Central Universities (No.WK2340000007), by One Hundred Person Project of CAS, by Shanghai Supercomputer Center and Supercomputing Center of USTC.

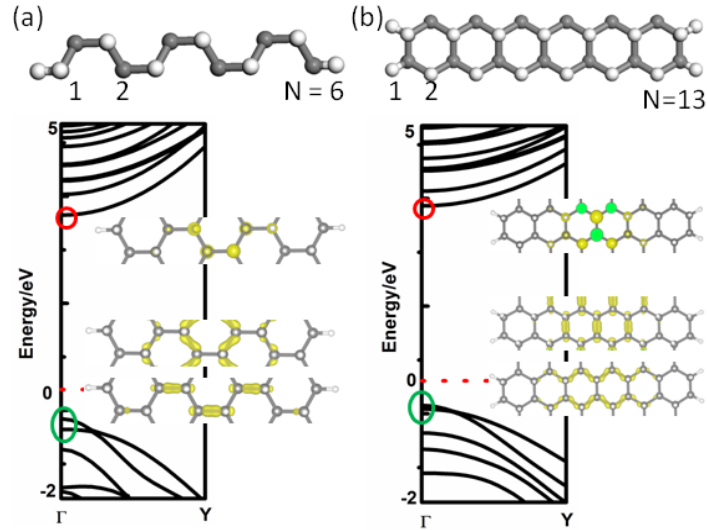

**FIG. 1 (Color online)** The optimized structures, electronic band structures and charge density profiles of CBE, VBE and the band just below the VBE (VBE-1) at Γ K-point are plotted for (a) 6-zigzag and (b) 13-armchair GANRs, respectively. The grey and white balls represent carbon and hydrogen atoms, respectively. The fermi energy level is set as zero, plotted with red dotted line.

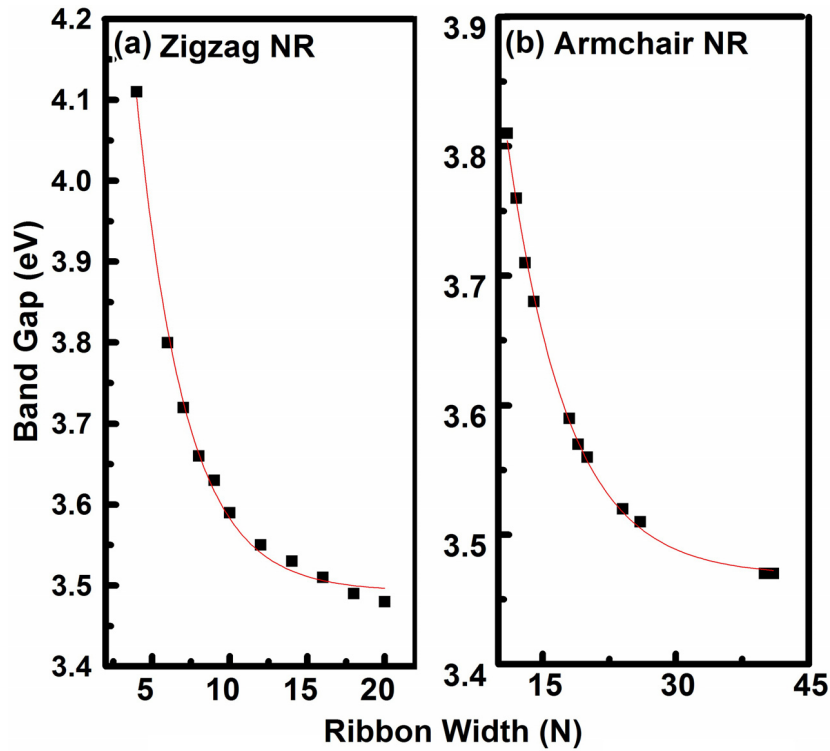

**FIG. 2 (Color online)** The energy band gaps of armchair and zigzag GANRs are plotted with the nanoribbons' widths.

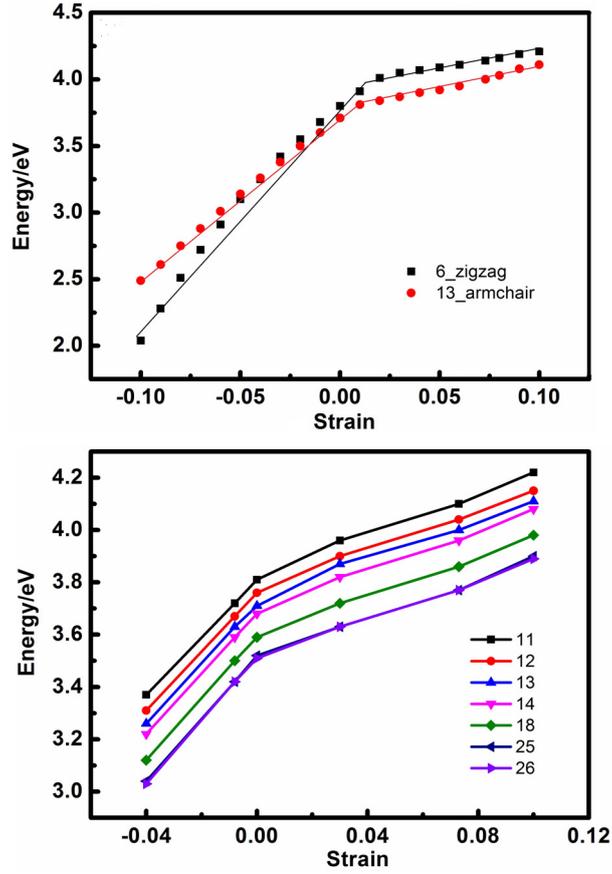

**FIG. 3 (Color online)** (a) The energy band gaps of 6-zigzag and 13-armchair GANRs are plotted with uniaxial strain. (b) The energy band gaps of armchair GANRs with various *N* (*N=11 to 26*) are plotted with uniaxial strain.

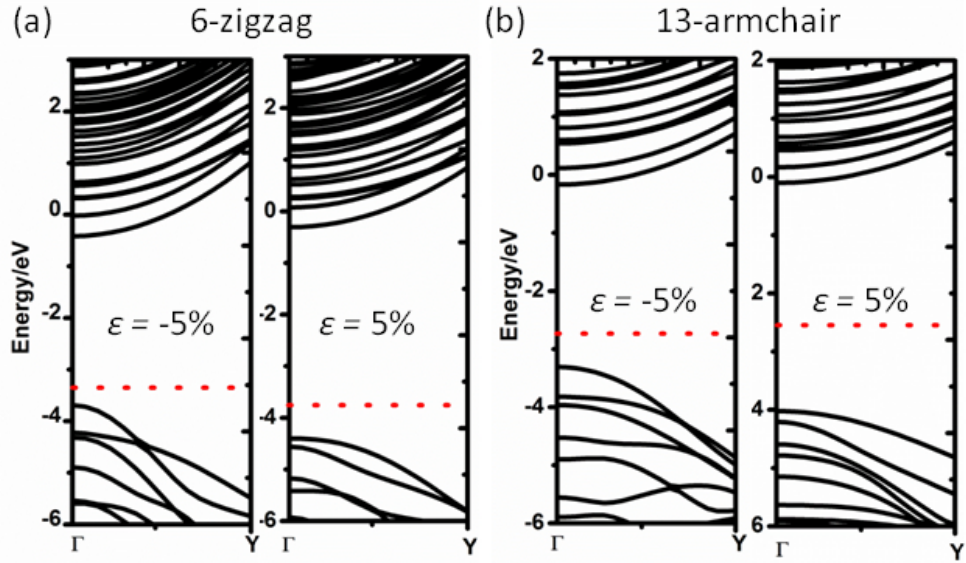

**FIG. 4 (Color online)** The electronic band structures of (a) 6-zigzag and (b) 13-armchair GANRs with and without uniaxial strain are plotted.